# The Geology and Geophysics of Kuiper Belt Object (486958) Arrokoth

J.R. Spencer[1]*, S.A. Stern[1], J.M Moore[2], H.A. Weaver[3], K.N. Singer[1], C.B Olkin[1], A.J. Verbiscer[4], W.B. McKinnon[5], J.Wm. Parker[1], R.A. Beyer[6,2], J.T. Keane[7], T.R. Lauer[8], S.B. Porter[1], O.L. White[6,2], B.J. Buratti[9], M.R. El-Maarry[10,11], C.M. Lisse[3], A.H. Parker[1], H.B. Throop[12], S.J. Robbins[1], O.M. Umurhan[2], R.P. Binzel[13], D.T. Britt[14], M.W. Buie[1], A.F. Cheng[3], D.P. Cruikshank[2], H.A. Elliott[15], G.R. Gladstone[15], W.M. Grundy[16,17], M.E. Hill[3], M. Horanyi[18], D.E. Jennings[19], J.J. Kavelaars[20], I.R. Linscott[21], D.J. McComas[22], R.L. McNutt Jr.[3], S. Protopapa[1], D.C. Reuter[19], P.M. Schenk[23], M.R. Showalter[6], L.A. Young[1], A.M. Zangari[1], A.Y. Abedin[20], C.B. Beddingfield[6], S.D. Benecchi[24], E. Bernardoni[18], C.J. Bierson[25], D. Borncamp[26], V.J. Bray[27], A.L. Chaikin[28], R.D. Dhingra[29], C. Fuentes[30], T. Fuse[31], P.L Gay[24], S.D.J. Gwyn[20], D.P. Hamilton[32], J.D. Hofgartner[9], M.J. Holman[33], A.D. Howard[34], C.J.A. Howett[1], H. Karoji[35], D.E. Kaufmann[1], M. Kinczyk[36], B.H. May[37], M. Mountain[38], M. Pätzold[39], J.M. Petit[40], M.R. Piquette[18], I.N. Reid[41], H.J. Reitsema[42], K.D. Runyon[3], S.S. Sheppard[43], J.A. Stansberry[41], T. Stryk[44], P. Tanga[45], D.J. Tholen[46], D.E. Trilling[17], L.H. Wasserman[16]

[1] Southwest Research Institute, Boulder, CO 80302, USA
[2] NASA Ames Research Center, Moffett Field, CA 94035-1000, USA
[3] Johns Hopkins University Applied Physics Laboratory, Laurel, MD 20723, USA
[4] Department of Astronomy, University of Virginia, Charlottesville, VA 22904, USA
[5] Department of Earth and Planetary Sciences and McDonnell Center for the Space Sciences, Washington University, St. Louis, MO 63130, USA
[6] SETI Institute, Mountain View, CA 94043, USA
[7] Division of Geological and Planetary Sciences, California Institute of Technology, Pasadena, CA 91125, USA
[8] National Science Foundation's National Optical Infrared Astronomy Research Laboratory, Tucson, AZ 26732, USA
[9] Jet Propulsion Laboratory, California Institute of Technology Pasadena, CA 91109, USA
[10] Department of Earth and Planetary Sciences, Birkbeck, University of London WC1E 7HX, London, UK
[11] University College London, Gower St, Bloomsbury, London WC1E 6BT, United Kingdom
[12] Independent Consultant, Washington, D.C., USA
[13] Massachusetts Institute of Technology, Cambridge, MA 02139, USA
[14] Department of Physics, University of Central Florida, Orlando, FL 32816, USA
[15] Southwest Research Institute, San Antonio, TX 78238, USA
[16] Lowell Observatory, Flagstaff, AZ 86001, USA
[17] Department of Astronomy and Planetary Science, Northern Arizona University, Flagstaff, AZ, 86011, USA
[18] Laboratory for Atmospheric and Space Physics, University of Colorado, Boulder, CO 80303, USA

[19] NASA Goddard Space Flight Center, Greenbelt, MD 20771, USA
[20] National Research Council of Canada, Victoria, BC V9E 2E7, Canada
[21] Independent Consultant, Mountain View, CA 94043, USA
[22] Department of Astrophysical Sciences, Princeton University, Princeton, NJ 08544, USA
[23] Lunar and Planetary Institute, Houston, TX 77058, USA
[24] Planetary Science Institute, Tucson, AZ 85719, USA
[25] Earth and Planetary Science Department, University of California, Santa Cruz, CA 95064, USA
[26] Decipher Technology Studios, Alexandria, VA 22314
[27] Lunar and Planetary Laboratory, University of Arizona, Tucson, AZ 85721, USA
[28] Independent Science Writer, Arlington, VT 05250, USA
[29] University of Idaho, Moscow, ID 83844, USA
[30] Universidad de Chile, Centro de Astrofísica y Tecnologías Afines, Santiago, Chile
[31] Kashima Space Technology Center, National Institute of Information and Communications Technology, Kashima, Ibaraki 314-8501, Japan
[32] Department of Astronomy, University of Maryland, College Park, MD 20742, USA
[33] Center for Astrophysics, Harvard-Smithsonian Center for Astrophysics, Cambridge, MA 02138, USA
[34] Department of Environmental Sciences, University of Virginia, Charlottesville, VA 22904, USA+D132
[35] National Institutes of Natural Sciences, Tokyo, Japan
[36] Marine, Earth, and Atmospheric Sciences, North Carolina State University, Raleigh, NC 27695, USA
[37] Independent Collaborator, Windlesham, England, GU20 6YW, UK
[38] Association of Universities for Research in Astronomy, Washington, DC 20004, USA
[39] Rheinisches Institut für Umweltforschung an der Universität zu Köln, Cologne 50931, Germany
[40] Institut Univers, Temps-fréquence, Interfaces, Nanostructures, Atmosphère et environnement, Molécules, Unité Mixte de Recherche, Centre National de la Recherche Scientifique, Universite Bourgogne Franche Comte, F-25000 Besancon, France
[41] Space Telescope Science Institute, Baltimore, MD 21218, USA
[42] Independent Consultant, Holland, MI 49424, USA
[43] Department of Terrestrial Magnetism, Carnegie Institution for Science, Washington, DC 20015, USA
[44] Roane State Community College, Oak Ridge, TN 37830, USA
[45] Université Côte d'Azur, Observatoire de la Côte d'Azur, Laboratoire Lagrange/ Centre National de la Recherche Scientifique, Unité Mixte de Recherche 7293, 06304 Nice Cedex 4, France
[46] Institute for Astronomy, University of Hawaii, Honolulu, HI 96822, USA

*Correspondence to: spencer@boulder.swri.edu.
2

**The Cold Classical Kuiper Belt, a class of small bodies in undisturbed orbits beyond Neptune, are primitive objects preserving information about Solar System formation. The New Horizons spacecraft flew past one of these objects, the 36 km long contact binary (486958) Arrokoth (2014 MU$_{69}$), in January 2019. Images from the flyby show that Arrokoth has no detectable rings, and no satellites (larger than 180 meters diameter) within a radius of 8000 km, and has a lightly-cratered smooth surface with complex geological features, unlike those on previously visited Solar System bodies. The density of impact craters indicates the surface dates from the formation of the Solar System. The two lobes of the contact binary have closely aligned poles and equators, constraining their accretion mechanism.**

On 2019 January 1 at 05:33:22 Universal Time (UT) the New Horizons spacecraft flew past the Kuiper Belt object (KBO) (486958) Arrokoth (2014 MU$_{69,}$ originally nicknamed "Ultima Thule"), at a distance of 3538 km (*1*). Arrokoth is a contact binary consisting of two distinct lobes, connected by a narrow neck. Based on its orbital semi-major axis, low eccentricity and inclination (*2*), and its albedo and color (*1,3*), Arrokoth is classified as a member of the dynamically cold, non-resonant "cold classical" KBO (CCKBO) population, and probably a member of the tight orbital clustering of CCKBOs known as the kernel (*4*). There is no known mechanism for transporting the majority of these objects onto these nearly circular orbits, so they are thought to have formed in situ and remained dynamically undisturbed since the formation of the Solar System. Due to the low impact rates (*5*) and low temperatures in the Kuiper Belt, CCKBOs are also thought to be physically primitive bodies. Arrokoth's equivalent spherical diameter of 18 km (see below), makes it about 8.5 times smaller in diameter than a known break in the size-frequency distribution of CCKBOs at diameter ~100 km (*6*).

Initial results from this flyby (*1*) were based on early data downlinked from the spacecraft. Since then, additional data have been downlinked, including: (i) the highest-resolution images from the flyby, taken with the narrow-angle Long-Range Reconnaissance Imager (LORRI) camera (*7*). These LORRI images have a pixel scale which is 4 times finer (33 m pixel$^{-1}$) than the 130 m pixel$^{-1}$ of previously-available Multicolor Visible Imaging Camera (MVIC) (*8*) images (*1*), though due to smear and a lower signal-to-noise ratio (SNR), the effective resolution of the LORRI images is only about 2 times better than the MVIC images; (ii) Additional LORRI images from earlier approach epochs, with higher SNR than previously-downlinked data; (iii) Improved LORRI distant approach rotational coverage, constraining the shape and rotational parameters; and (iv) Additional satellite and ring search data from LORRI and MVIC. See (*9*) for image processing details. We describe Arrokoth's shape, geological evolution, and satellite and ring constraints resulting from these additional data, and from continued analysis of all downlinked data.

**Stereo Imaging**

A pair of LORRI images designated CA04 and CA06 (Fig. 1A, Table S1 (*9*)) provides improved stereo imaging to constrain the shape and topography of the close approach hemispheres of the two lobes. A stereographic terrain model derived from these images, (Data S1, (*9*)), is shown in Fig. 2. Topographic relief in the stereo model is ~0.5 km or less on both lobes (away from the neck region), similar to the 1.0 km and 0.5 km relief seen in limb profiles of the large and small lobes respectively (*1*). The stereo images (Fig. 1A) show additional topographic detail that is visible by eye but is below the 200 meter vertical resolution of the terrain model. Our interpretation is based on both the terrain model and subjective analysis of the stereo pair.

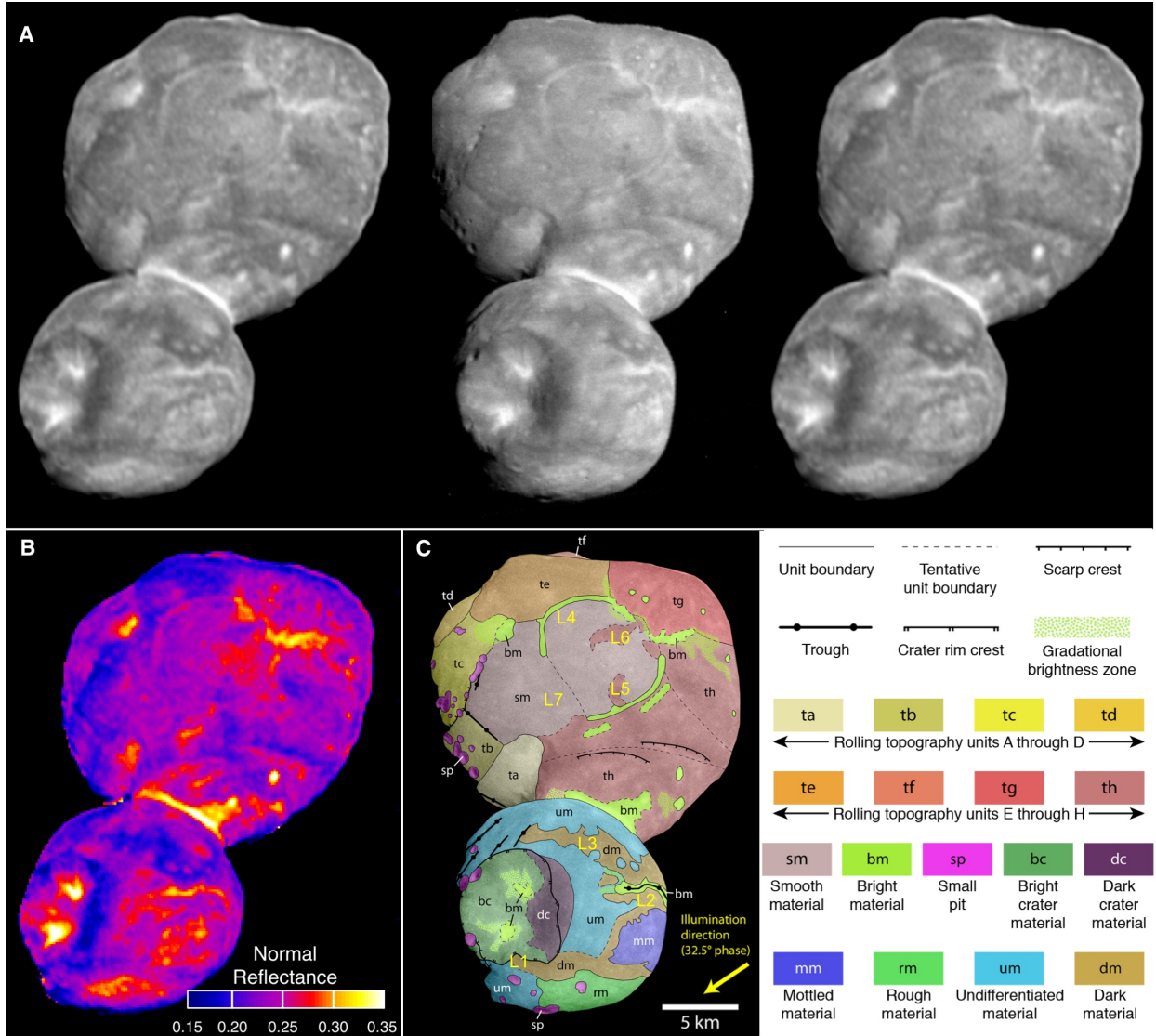

**Fig. 1**. **Mapping of Arrokoth.** A. Cross-eyed (left+center) and direct (center+right) stereo pair image of Arrokoth, taken by LORRI. The left and right images are CA04, range = 27,860 km, phase = 12.9°, 138 m pixel$^{-1}$, while the center image is CA06, range = 6,650 km, phase = 32.5°, 33 m pixel$^{-1}$. Both images have been deconvolved to remove the LORRI point-spread function, and motion blur from CA06, to maximize detail (*9*). B. 0.6 µm normal reflectance map of Arrokoth, based on image CA04. C. Geomorphological map of Arrokoth, overlain on the deconvolved CA06 image. The positive spin axis of Arrokoth is pointing approximately into the page. Yellow labels L1 – L7 identify locations mentioned in the text. Geological units are labelled and colored as shown in the legend.

**Rotation and Global Shape Modeling**

No periodic brightness variation due to rotation was detected in Hubble Space Telescope (HST) photometry before the flyby, with an upper limit amplitude of about 0.15 magnitudes (*10*). Stellar occultations in July 2017 and August 2018 showed that Arrokoth had an elongated, possibly contact-binary shape (*11*). The elongated shape and the low lightcurve amplitude implied that Arrokoth's rotational pole was roughly aligned with the direction of the Sun and Earth.

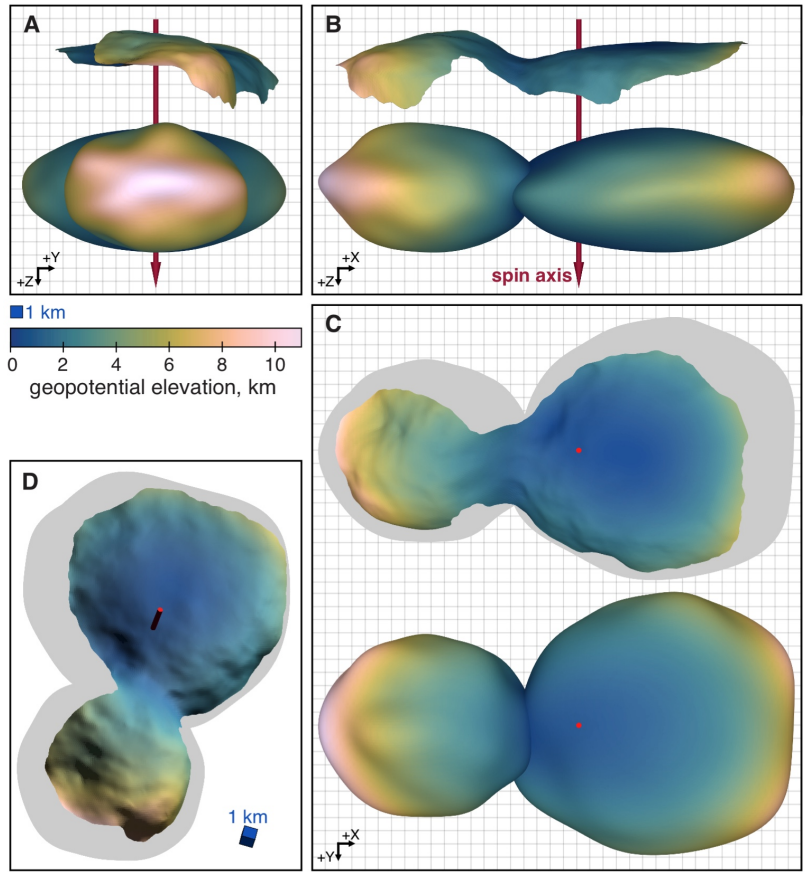

**Fig. 2. Stereo and global shape models.** A – C: Comparison of the stereo shape model of the encounter face (top of each panel) to the global shape model (bottom of each panel), as seen from the -X (small lobe) direction (A), the +Y direction (B), and the south polar (-Z) direction (C). The red arrow shows the orientation and location of the positive spin axis. Each model is colored to show the variation in geopotential across the surface. The stereo model has been trimmed to remove edge effects. D: Stereo model seen from the same geometry as the CA06 observation (Fig. 1A, center), but with different lighting, chosen to highlight the small-scale topography.

Arrokoth's rotation and global shape are mostly determined from LORRI images taken between 2.2 days before the encounter, when Arrokoth first exceeded 2 pixels in length, and 9 minutes after encounter, when Arrokoth was last imaged (at high phase angle) as a receding crescent (Fig. 3). Disk-integrated photometry from earlier unresolved LORRI images showed no periodic variations in brightness, with an upper limit amplitude of 0.1 magnitudes (*12*), but were affected by confusion from the dense stellar background. The strongest constraints on the shape model are from a series of approach images with cadence between 1 hour and 20 minutes, starting 13.6 hours before closest approach, when Arrokoth subtended 10 pixels in length (Fig. 4A). These images covered 85% of the 15.92-hour rotation period, though only one hemisphere of Arrokoth was visible because of the near-alignment of the rotational pole with both the direction of the Sun and New Horizons' approach direction.

Incorporating the additional rotational coverage images now available into the same rotational modelling techniques as before (*1*), the rotational period of Arrokoth is unchanged at 15.92 ± 0.02 hours, but its pole orientation has been refined. The positive rotational pole points to Right Ascension 317.5 ± 1°, Declination -24.9 ± 1° in the J2000 equinox. The rotation rate is within the range of other CCKBOs (*13,14,15*). The resulting obliquity of Arrokoth's pole to its orbit is 99 ± 1°, and the rotational pole is 39 ± 1° from the New Horizons approach vector and 28 ± 1° from the direction from the Sun to Arrokoth during the encounter. The rotational brightness variation implied by the shape model would have a peak-to-peak amplitude of 0.05 magnitudes from New Horizons' approach direction, consistent with the earlier non-detections.

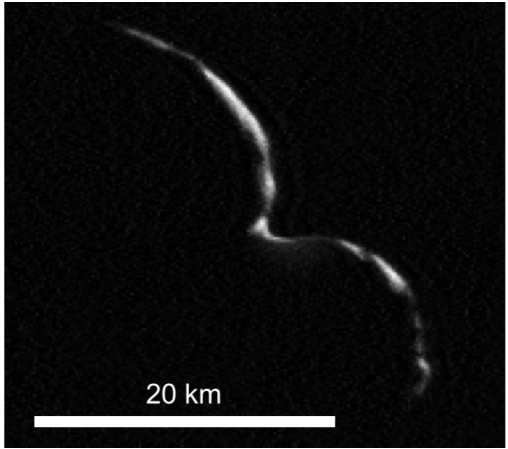

**Fig. 3. Arrokoth seen at high phase.** New Horizons' last view of Arrokoth (CA07), taken with the LORRI camera 9.4 minutes after closest approach at phase angle 152º, range 8,800 km, and resolution 175 m pixel$^{-1}$. This image has been deconvolved to remove the motion smear visible in Fig. 4B (*9*). The large lobe is in the upper left and

A low-resolution global shape model (Data S2 (*9*)) was produced using all available observations—even the early, distant ones—to refine the model. The high phase angle CA07 observation (Fig. 3, Table S1, (*9*)), of the illuminated double-crescent of Arrokoth, provides a constraint on how thick the unilluminated side can be, based on which stars are and are not eclipsed by the object (Fig. 4B). There remain differences between the shape model and the LORRI images in Fig. 4A, e.g. compared to the model, the images show a less indented neck and flatter outer end of the small lobe between December 31 20:38 and January 1 01:12.

The best-fitting global shape model consists of two roughly ellipsoidal lobes with overall dimensions X, Y, and Z of 36 × 20 × 10 km. Maximum dimensions of the large and small lobes are 20.6 × 19.9 × 9.4 km and 15.4 × 13.8 × 9.8 km respectively. The uncertainty for these dimensions is roughly 0.5 × 0.5 × 2.0 km in X, Y, and Z respectively; larger in the Z direction because the flyby imaged little of the +Z (northern) half of the object. The total volume is 3210 ± 650 km$^3$, equivalent to a sphere of diameter 18.3 ± 1.2 km. This volume is 30% larger than the previous estimate of 2450 ± 720 km$^3$ (*1*), though consistent within the uncertainties. The larger lobe has a volume equal to a sphere of diameter 15.9 ± 1.0 km, while the equivalent diameter for the smaller lobe is 12.9 ± 0.8 km. These values give a volume ratio (and mass ratio if densities are equal) of 1.9 ± 0.5.

Fig. 2 compares the global shape model to the stereo model of the encounter (-Z) side of Arrokoth. There is broad agreement between the two techniques, though the south polar region of the large lobe is flatter in the stereo model, and the neck is smoother (a slope discontinuity at the neck is an intrinsic feature of the global shape model, due to its dual-lobe nature). We regard the stereo model as more reliable than the global shape model in the south polar and neck regions, because the stereo model incorporates additional information due to the matching of albedo features, and because these albedo features can also produce artifacts in the global shape model, which assumes a uniform surface albedo. However, near the limbs the stereo model performs poorly because foreshortening makes feature matching difficult, while the global shape model is well constrained near the limbs.

**Gravity Modeling**

The irregular shape of Arrokoth produces a complex geophysical environment. We calculated Arrokoth's geopotential (the sum of the gravitational and rotational potentials in a body-fixed reference frame) using the low-resolution global shape model, the 15.92-hour rotation period, and an assumed bulk density. In the absence of spacecraft gravity measurements or detected satellites, the density of Arrokoth is not directly constrained. However, if the neck of Arrokoth is assumed to have no tensile strength, the density must be >290 kg m$^{-3}$, or the rotation would overcome the mutual gravity of the two lobes, causing them to separate. We assume a nominal bulk density of 500 kg m$^{-3}$, similar to the measured densities of cometary nuclei (e.g. comet 67P/Churyumov-Gerasimenko (*16*)), which leads to a mean surface gravity of ~1 mm s$^{-2}$. If this density is correct, the requirement for the two lobes to support each other against their mutual gravity over their ~28 km$^2$ contact area, implies a compressive strength (accounting for centrifugal force) of > 2.3 kPa.

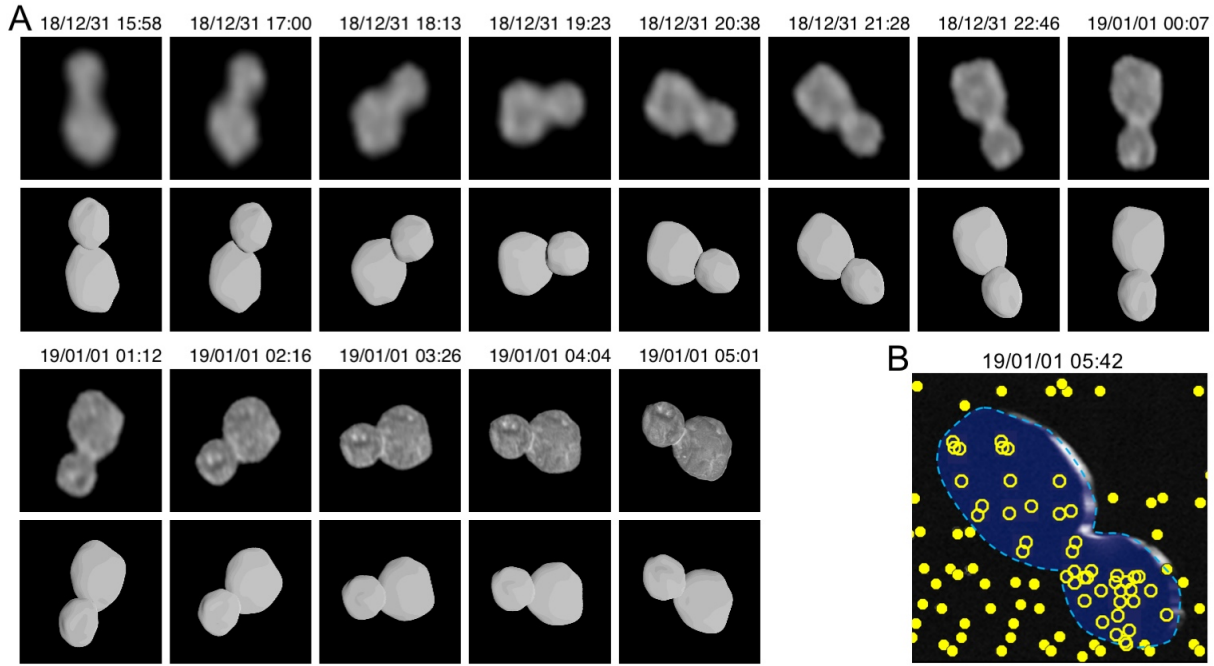

**Fig. 4. Shape model compared to LORRI images**. A: Deconvolved LORRI approach images of Arrokoth, compared to synthetic images with the same geometry derived from the global shape model. Images have been scaled to a constant frame size of 44 x 44 km, so become sharper as time progresses and range decreases. Celestial north is up. B: the CA07 departure image, with the silhouette (dark blue) and outline (light blue dashed line) of the shape model superposed. Open and filled yellow dots indicate the locations of occulted and unocculted stars respectively in the 6-frame CA07 sequence, used to constrain the shape of the unilluminated hemisphere.

Figure 2 uses color to show the geopotential altitude, calculated by dividing the geopotential by the total acceleration, which represents elevation with respect to a gravitational equipotential surface (*17*). The geopotential is calculated from the global shape model, then evaluated on the surfaces of the global shape model and the stereo model (with positions matched to the global shape model (*9*)). This approach results in slight inaccuracies in the geopotential calculated across the stereo model, as there are regions where the stereo model rises above/below the surface of the global shape model. We focus on general trends that are robust to the uncertainties in the shape model. The geopotential is highest at the distal ends and equator, and decreases with increasing latitude on each lobe, reaching a global minimum at the neck. For an assumed density of 500 kg m$^{-3}$, surface slopes (derivatives of the geopotential (*17*)) are generally gentle (<20°) and slope downward to higher latitudes and into the neck region (Fig. S1). If material can flow downslope, then it will collect at higher latitudes and in the neck region. The stereo model shows that the neck is relatively smooth compared to its sharp appearance in the global shape model, with shallow slopes. The global shape model shows slopes of >30° at the neck, but this steepness is in part an artifact of the global model's treatment of Arrokoth as two separate overlapping bodies.

The configuration of the two lobes of Arrokoth has implications for its formation and evolution (*1,18*). Using the same assumptions as above, we calculate the principal axes of inertia for the two lobes by slicing Arrokoth's neck at the narrowest point. This confirms that the large lobe's highest moment of inertia axis is aligned within < 5º of its small lobe counterpart, and the equatorial planes of the two bodies are also almost coincident in space, with the estimated center of mass of the small lobe displaced only 0.2 km from the equatorial plane of the large lobe.

**Surface Units**

Fig. 1B shows a map of 0.6 µm normal reflectance (*19*). The map is derived from the high-SNR CA04 image, using a merger of the global and stereo shape models to determine illumination at each point, and an assumed lunar-like photometric function which has no limb darkening at zero phase (*20*). The normal reflectance is equal to the geometric albedo of a body covered in material with that location's photometric properties. Arrokoth's mean 0.6 µm normal reflectance, and thus its geometric albedo, is 0.23. The mean and standard deviation of the normal reflectance are 0.230 and 0.035 respectively for the large lobe, and 0.228 and 0.043 respectively for the small lobe.

We have also produced (*9*) an updated geological unit map of Arrokoth (Fig. 1C) that supersedes the previous preliminary map (*1*). Note that this mapping is physiographic in nature and is not intended to rigorously convey stratigraphic relations between units. The small and large lobes have distinctly different surface appearances, so we mapped their surface units separately and describe them separately below.

**Small Lobe**: This lobe is dominated by a large depression (informally named Maryland), which is very likely to be an impact crater (*1*). The projected crater rim measures ~6.7 by 6 km across in the image plane, with its longer axis roughly aligned with the principal axis of Arrokoth. The ellipticity might be due to foreshortening, in which case Maryland could be circular with diameter 6.7 km. Stereo measurements show that the deepest well-determined point in Maryland is 0.51 km below a plane defined by the rim, or 1.3 km below the surface of a sphere with the small lobe's mean radius, giving a depth/diameter ratio of 0.08 – 0.19. This depth/diameter ratio is similar to craters on other bodies with gravities similar to Arrokoth's ~1 mm s$^{-2}$, including asteroids Šteins (~0.12, 0.8-1.3 mm s$^{-2}$, (*21*)) and Eros (~0.13, 2.4-5.5 mm s$^{-2}$, (*22*)), though these bodies are composed of different materials and may have different porosities. Stereo imaging (Fig. 1A) reveals that the part of its rim furthest from the large lobe features a promontory protruding into the crater (marked L1 in Fig 1C), at an elevation similar to the rest of the rim, which is not a common feature of impact crater rims.

Albedo patterns across the small lobe are complex. There are two patches of bright material (unit *bm*) within Maryland, which show discrete boundaries near the crater bottom, and fade towards the crater rim. Straddling the Maryland rim on the side opposite the bright patches is discrete, dark crater rim material (unit *dc*), which contrasts with the brighter terrain (unit *bc*) that forms the remainder of the crater interior. Elsewhere on the small lobe, discrete morphological units have albedo variations of almost a factor of two (Fig. 1B). The rough terrain at the distal end of the small lobe (unit *rm*), forms a facet that is relatively flat compared to the overall curvature of the surface, and is brighter than its immediate surroundings. The low illumination angle on this facet reveals a rough surface texture at a scale of a few hundred meters, apparently mostly composed of sub-km pits, with one prominent ~150 m diameter pit (marked 27 in Fig. 6A) which resembles a small, fresh, bowl-shaped impact crater. Another nearby mottled bright unit (*mm*), may be similar, but is seen at a higher illumination angle so topographic roughness is not apparent, and has a distinctly crenulated and angular margin relative to that of unit *rm* (L2 in Fig. 1C).

Dark material surrounding the *mm* unit seems to be part of a discrete unit, designated *dm*, that wraps around much of the remainder of the observable surface of the small lobe - this material is the darkest on Arrokoth, with minimum 0.6 µm reflectance of 0.18. In places (L3 in Fig. 1C), it has a boundary with pointed and angular protrusions and rounded indentations, which may indicate material erosion and removal due to scarp retreat (*1*). Near L3 in Fig. 1C, there are also bright circular patches within the dark material. Running down the center of the principal mapped outcrop of dark material is a sinuous unit of bright material (unit *bm*), which stereo observations show occupies a V-shaped trough. The rest of the surface of the small lobe is nondescript at the available lighting and resolution and has been mapped as undifferentiated material (unit *um*). Crossing the undifferentiated material near the terminator between Maryland and the large lobe are a series of roughly parallel troughs, which are reminiscent of structural troughs seen on other similar-sized bodies, for instance asteroid Eros (*23,24*), Saturn satellites Epimetheus and Pandora (*25*), and the Martian satellite Phobos (*26*).

Our data confirm that the bright "neck" region connecting the two lobes has a diffuse margin at least on the large lobe side, but extreme foreshortening makes it difficult to characterize its margin on the small lobe side.

**Large Lobe**: The larger lobe is very different in appearance from the small lobe. Previous analysis (*1*) mapped the large lobe as composed of a series of roughly equally-sized, discretely bounded, rolling topographic units. We interpret some of these units and their boundaries differently, though confirm the discrete nature of many of the units (*ta* through *tg*). Those near the terminator, *ta – td*, are distinctive, being brighter than adjacent units (Fig. 1B) (though *ta* is noticeably less red than the others (*3*)), and are clearly separated from the rest of the large lobe by a common, continuous scarp or trough and chain of pits. Units *tg* and *th* appear more mottled than adjacent units, and stereo imaging of these suggests that their surface consists of dark ridges and hills surrounded by brighter low terrain.

The rest of the large lobe is occupied by smooth material (unit *sm*) of moderate albedo, transected by a series of distinctive bright linear features (unit *bm*), some of which form an incomplete annulus. In some areas (e.g., L4 in Fig. 1C) the inner margin of the annulus appears sharply bounded, possibly with an outward-facing scarp, while the outer margin is more diffuse. Stereo observations (Figs. 1A, 2D) show that terrain within the annulus is flatter than the undulating surface of the rest of the visible portion of the large lobe, and suggest that the annulus occupies a shallow trough. At the boundary between units *tg* and *sm*, the annulus appears to be interrupted by diffuse bright material, which may be superimposed upon it. In two places, L5 and L6 in Fig 1C, dark hills appear to extend into the *sm* unit. At L5 in Fig. 1C, these hills seem to be an extension, cut by the *bm* annulus, of similar hills on unit *th*. We discuss the possible origin of these features below.

**Geological Interpretation**
Our data, particularly the stereo images, confirm that the brighter material on both lobes occurs preferentially in depressions. The brightest material on the large lobe (the possible crater numbered 17 in Fig 6A), on the small lobe (bright features 42 and 43 in Fig. 6A), and in the bright collar between the two lobes all have normal 0.6 µm reflectance values near 0.37, suggesting that the bright material has similar chemical and physical properties in all these regions. The most extensive bright region, the bright collar in the topographic low of the "neck" region, may be simply the largest-scale example of a general process that creates bright low-lying material across Arrokoth. As proposed in (*1*), loose, poorly consolidated, likely fine-grained bright material may move downslope and accumulate in depressions, which would imply that bright material is more mobile than dark material on Arrokoth. The complex albedo patterns on the small lobe, and their crenulated margins, may result from the exposure and differential erosion of multiple lighter and darker layers oriented roughly parallel to its surface, though independent topographic information is of insufficient quality to confirm this explanation.

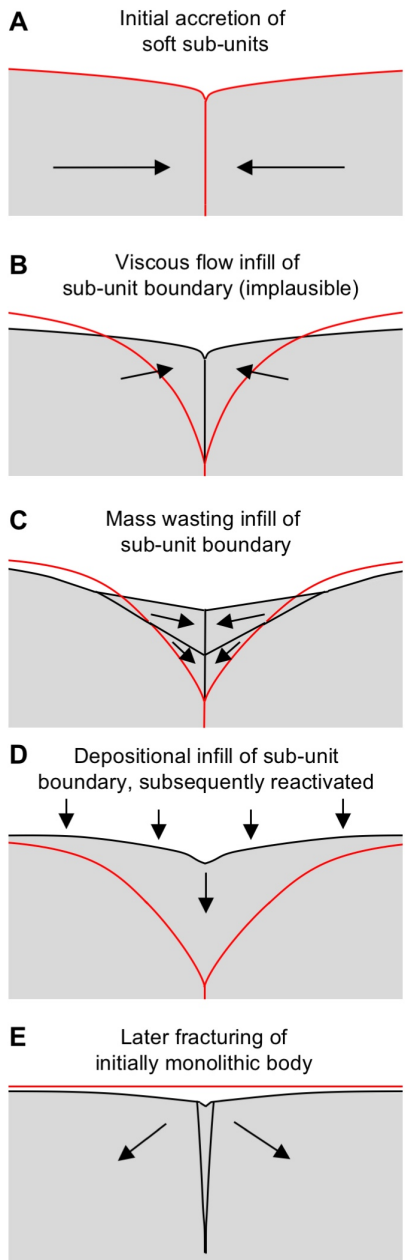

**Fig. 5. Possible explanations for the appearance of the boundaries between terrain sub-units on the large lobe.** The original surface (shown in red) is modified by the processes labelled in each panel. We consider options D and E to be most consistent with the available evidence; see text for discussion.

It was previously proposed (*1*) that the large lobe might be composed of smaller sub-units that accreted separately. However, the improved imagery and topography raise issues with this interpretation. Firstly, the central *bm* annulus, enclosing what was mapped as a discrete sub-unit in (*1*) appears to be younger than some other surface features, and not an unmodified primordial boundary, for the following reasons: (i) the annulus is incomplete, with no discernable topographic feature or textural change in the gap region where it is missing (L7 in Fig 1C)- for this reason we map a continuous unit, *sm*, across this gap; (ii) even where the annulus is conspicuous, it cuts across flat terrain for most of its length; and (iii) dark hills found on the *th* and *sm* sub-units appear to form a continuous physiographic unit cut by the annulus (at L5 in Fig 1C), and (iv) the partially concentric nature of the annulus suggests a structural basis, not greatly obscured by subsequent deposition. Secondly, though other proposed sub-units are distinguishable by differing surface textures, albedos and modest topographic inflections or other surface features, the overall shape of the large lobe is smooth and undulating. There are no major topographic discontinuities between the sub-units comparable to that between the two lobes, as would be expected if the sub-units had similar internal strength to the lobes as a whole. Erosion and alteration over the past 4.5 Ga (see below) is likely to have modified the optical surface and the uppermost few meters (*27*) but probably does not explain the smoothness seen at the > 30 meter scale of the New Horizons imaging resolution.

Some possible explanations for the appearance of the annulus and other sub-unit boundaries are illustrated in Fig. 5. The sub-units may have been soft enough at the time of merger that they conformed to each other's shapes on contact (*28, 29,1*) (Fig 5A), though no evidence for impact deformation is seen. In order for such deformation to take place at the time, the shear strength of the merging components must have been no more than 2 kPa, the ram pressure of an impacting body assuming a merger velocity of 1-2 m s$^{-1}$ and a material density of 500 kg m$^{-3}$. The possibility that sub-units flowed viscously due to gravity after contact while still soft (Fig. 5B) can be discounted, because such flow would require an implausibly low shear strength of ~100 Pa. Erosion and downslope movement (mass wasting) may have filled in original gaps between the sub-units (Fig. 5C), though there is an absence of obvious boundaries (except perhaps at the the *tg/sm* contact) between material transported by mass wasting and in-situ material. The fact that mass wasting has not filled the much larger depression between the two lobes also implies that any major mass wasting process must have ceased before the merger of the two lobes. The original discontinuities may have been buried by subsequent accretion or redistribution of surface material (Fig. 5D). The boundaries would then need to be re-activated in some way to still be visible on the surface, possibly by collapse into subsurface voids or degassing of volatiles such as $N_2$ or CO, which may explain the trough-like appearance of parts of the *bm* annulus, and the troughs and pit chains seen at low illumination angle between the *ta – td* sub-units and the rest of the larger lobe. However, it's not clear how burial could preserve different surface textures for the different sub-units. Alternatively, the large lobe may be monolithic, and the visible boundaries may be secondary features (Fig. 5E), e.g. produced by subsequent fracturing. For the annulus, we consider the evidence to be most consistent with scenarios D and E in Fig. 5. However, in any of these cases, the processes that produced the distinctive surface textural contrasts between the units, in particular the patches of dark hills and ridges, are unknown.

**Pits and Craters**

In addition to the 7 km diameter probable impact crater Maryland, scattered across the body of Arrokoth are numerous roughly circular sub-km bright patches and pits, though even if these are mostly impact craters the crater density is relatively low compared to many other small bodies (*1*, Fig. S2). The bright patches are generally seen in areas that have high illumination angle and are away from the terminator. Some of these patches appear in stereo imaging (Fig. 1A) to occupy depressions. These may be equivalent to the pits seen in low illumination angle areas near the terminator (unit *sp*, Fig.1C): these pits might also feature bright material on their floors that is invisible due to the unfavorable lighting.

We have classified these bright patches and pits to reflect our confidence that they are impact craters, based on the morphology expected for either fresh or degraded impact craters (*9*, Supplementary Text), as determined by multiple independent investigators. Crater candidates and their classifications are listed in Data S3 and shown in Fig. 6A. Our criteria included the spatial arrangement of the potential craters and their relationship to other geologic features. For instance, as noted above, a chain of pits that is coincident with a scarp on the boundary between units *tc* and *sm* possibly originated via surface collapse rather than impact (*1*). For a fresh crater formed on a flat and smooth surface, a crater rim is expected to be close to circular and raised above the surrounding terrain (unless the terrain is substantially porous (*30*)), though image resolution does not always allow identification of a raised rim. The interior shape of a crater is expected to be bowl-like with a depth/diameter ratio typically not higher than ~0.2 (*31*). The predicted modal impact velocity onto Arrokoth is ~300 m s$^{-1}$ (*5*), which is sufficient to form craters with typical morphologies (see Supplementary Text). In the case of Arrokoth, the lowest velocity impacts (≲ 20 m s$^{-1}$) are unlikely to leave conspicuous depressions, but these impacts are expected to be a small fraction of the total (*5*). The formation of a crater on a slope or modification by later geologic processes (such as mass wasting or a subsequent fault near the crater) may also alter the crater's appearance.

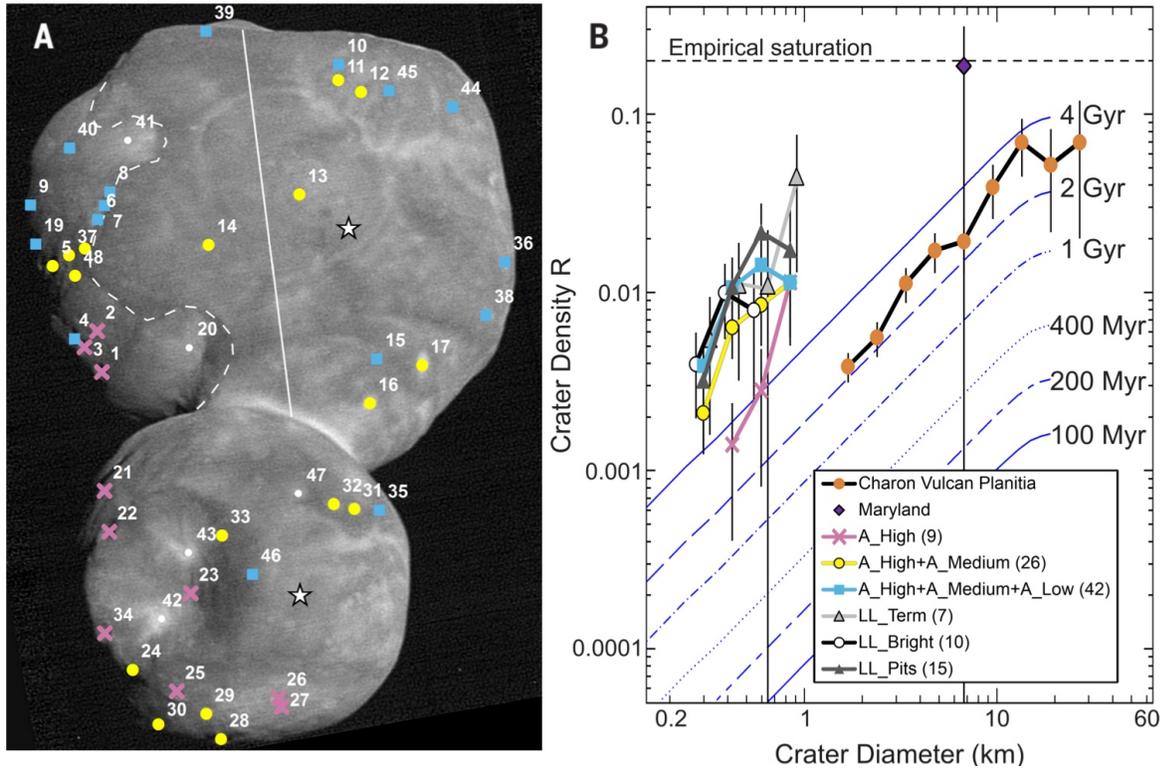

**Fig. 6**. **Craters and Pits on Arrokoth.** A. Locations of features considered for crater analysis: numbers refer to crater listings in Data S3. Color denotes confidence class: pink = high confidence (A_High), yellow = medium confidence (A_Medium), light blue = low confidence (A_Low). Features indicated in white are considered to be highly unlikely to be of impact origin and are not included in the crater statistics. The solid white line splits the large lobe into regions with differing lighting conditions, a more obliquely-illuminated region with more visible depressions (LL_Pits, left) and a more vertically-illuminated region with bright spots (LL_Bright, right). The white dashed curve delineates the boundary of combined geologic units ta, td, tc, and td, (LL_Term), considered together for crater density determination. The star symbols indicate the planetocentric subsolar point on each lobe according to the shape model. Lighting direction is shown in Fig 1C. B. The size-frequency distribution of craters on Arrokoth for each crater subgroup and region described in the text and shown in panel A, and (*9*). The yellow curve includes both high and medium confidence classes, and the light blue curve includes all confidence classes. Parenthetical numbers are the total number of craters/pits in each category. The Arrokoth crater data are compared to crater densities on Charon's Vulcan Planitia (*39*) without diameter adjustments for gravity or velocity scaling, and to predictions based on an impactor flux model for six different ages of surfaces on Arrokoth and gravity regime scaling (blue curves with different line styles, (*5*)). The LL_Term and LL_Bright distributions are offset horizontally by ± 9% for clarity. The empirical saturation line refers to a D$^{-3}$ differential power law distribution (*72*).

Potential small craters were subdivided in three ways (see Fig. 6A, (*9*)): (i) all pits and bright patches were subdivided based on our confidence that they are impact craters, (ii) features on the large lobe were subdivided into pits nearer the terminator, and bright patches away from the terminator, as shown in Fig. 6A, and (iii) a combination of geologic units ta, tb, tc, and td, designated "LL_Term" as they are on the large lobe terminator (Fig. 6A), was analyzed separately, because the entire combined unit has low-angle lighting optimal for crater identification. These subdivisions yielded a range of plausible crater densities, shown in Fig. 6B as a crater relative- or R-plot (*9*). Overall R values for each dataset are somewhat uncertain as they depend on the areas used for each distribution, and densities are lower if uncertain craters are excluded. The resulting uncertainty range of crater densities is less than a factor of 10 in each diameter bin in Fig. 6B.

Besides Maryland, all other possible impact features are 1 km in diameter or smaller. While the diameter gap between Maryland and second-largest crater on Arrokoth is large, the gap does not strongly disfavor a single power-law size distribution for the craters. We tested a model crater population with a power-law size distribution with slope $q = -2$ against the observed Arrokoth craters in the combined "A_High" and "A_Medium" categories. The resulting Anderson-Darling statistic indicates no substantial disagreement between the model and observed sample, with a significance level of $p \leq 17\%$.

Our analysis shows that Arrokoth appears to be only modestly cratered, relative to heavily cratered small objects like Phobos (Fig. S2), and there are some areas on Arrokoth where very few, if any, potential craters exist, in particular the part of the large lobe between the dashed and solid white lines in Fig. 6A.

The age of the surface can be estimated from the observed crater density. We converted impact flux estimates for Arrokoth to crater densities corresponding to several surface ages (*5*) and show these in Fig. 6B. The resulting age estimates are uncertain, given the uncertainty in identifying which craters are impact-generated, and because the model curves shift based on the crater scaling parameters used. Scaling in the strength regime, as opposed to scaling in the gravity regime assumed here (*5*), could in principle reduce the sizes of craters produced, if the surface strength of Arrokoth were sufficiently high. The expected strengths of porous cometary surfaces are, however, generally low enough (~1 kPa or less (*32*)) that the observed craters on Arrokoth should have formed in the gravity regime. In contrast, accounting for the additional cratering in an early but brief dynamical instability phase in the outer Solar System (*33*) would shift the model curves in Fig. 6B upward, although possibly by no more than a factor of two (*5*). Low relative densities of small craters are also observed on near-Earth asteroids, and are conventionally explained as due to seismic shaking from larger impacts or surface evolution due to changes in spin state (*34,35,36*). However, Arrokoth's spin state is likely to have evolved only very slowly (*18*), there do not appear to be sufficient impacts to act as effective seismic sources, and Arrokoth's likely high porosity would make seismic energy propagation highly inefficient. Overall, despite the paucity of craters on its surface, the observed crater density is consistent with a crater retention age of greater than ~4 billion years. The visible surface at the scale of the LORRI image resolution thus plausibly dates from the end of Solar System accretion.

Though the diameters of observed craters on Arrokoth (apart from Maryland) are smaller than those measured in the Pluto system, the slopes of the Arrokoth and Pluto system craters are consistent given the small number statistics. Using the Approximate Bayesian Computation forward-modeling methods (*37*, *38*), we estimated the posterior probability density functions for the parameters of independent truncated power-law crater size-frequency distribution models for Arrokoth's and Charon's (*39*) observed crater populations (for craters < 10 km diameter, below the break in slope observed on Charon). We then conducted the same analysis for a model with a common slope q between the two populations, but a separate offset. The mean slope $q = -1.8^{+0.4}_{-0.6}$ for Charon alone, $q = -2.3^{+0.6}_{-0.6}$ for Arrokoth alone, and $q = -2.0^{+0.4}_{-0.3}$ for the joint set (95% confidence). However, as seen in Fig. 6B, crater density on Arrokoth is higher than would be obtained from an extrapolation of the Charon slope and density to sub-km craters.

## Satellites and Rings

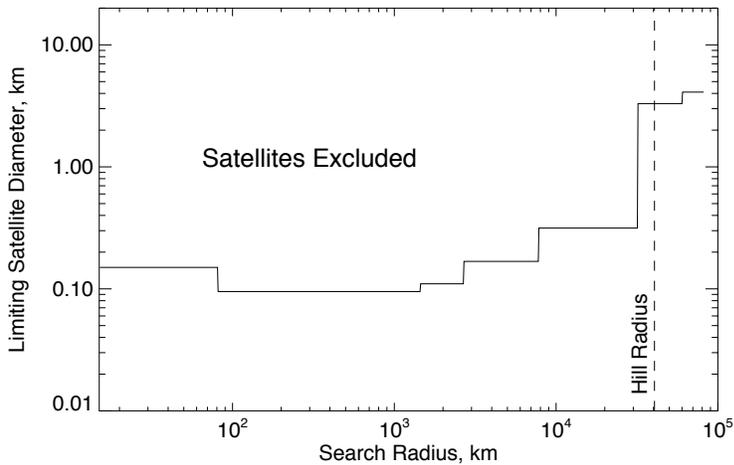

**Fig. 7. Upper limits on possible satellites of Arrokoth.** Excluded regions are plotted as a function of radius from the primary center of mass. The limits assume a satellite with photometric properties similar to Arrokoth itself. Gravitationally bound objects must lie within the Hill radius (dashed line), which is calculated assuming Arrokoth has a density of 500 kg m$^{-3}$.

Prior to the Arrokoth flyby, constraints on the prevalence of satellites and rings around sub-100 km diameter Kuiper Belt objects were limited. Larger CCKBOs are frequently members of orbiting binary pairs (*40*). Satellites with a primary/secondary brightness ratio larger than 20 have not been found for KBOs smaller than 500 km diameter (*41*), though this is likely in part due to observational biases. In contrast, satellites with high primary/secondary brightness ratio are common around large KBOs in non-CCKBO populations. The presence or absence of satellites provides a constraint on formation of the Arrokoth contact binary (e.g. a satellite could potentially remove angular momentum from the central body). At least two known asteroid contact binaries have small satellites: the large Trojan asteroid Hektor has a satellite which orbits at only 5 times the primary radius and has a diameter of 5% of the primary (*42*), and the large bi-lobed main-belt asteroid Kleopatra has two known satellites orbiting at 8 and 12 times the primary radius, with diameters 6% that of the primary (*43*).

New Horizons conducted a nested series of satellite searches with the LORRI camera during its approach to Arrokoth, using stacks of many images taken using 4x4 pixel binning to increase sensitivity and reduce data volume. Our dataset allows a deeper and broader search than previously reported *(1,9)*. No satellites have been found. We can exclude satellites larger than 100 – 180 meters in diameter (~0.5% the diameter of the primary) on orbits ranging from Arrokoth's surface to 8000 km radius, and < 300 m diameter throughout most of the Hill sphere (the region within which a moon could be gravitationally bound to Arrokoth), assuming albedos similar to Arrokoth itself (Fig. 7). Satellites analogous to those of Hektor and Kleopatra can thus be excluded.

The prevalence of rings around small KBOs is poorly constrained, but they are known around Chariklo (*44*), Haumea (*45*), and perhaps Chiron (*46*). We searched for rings and dust clouds within the Arrokoth environment at all phases of the encounter. The LORRI satellite searches on approach, discussed above, constrained backscattered light due to any ring or dust clouds to I/F $\lesssim 2 \times 10^{-7}$ (*19*) at 11º phase for a 10-km-wide ring, assuming neutral colors (*1*). This limit is fainter than Jupiter's main ring (I/F = 7 x $10^{-7}$ at 11º phase, (*47*)). We also conducted dedicated ring searches in forward-scattered light after closest approach, using images taken 1.7 – 2.3 hours after closest approach at a phase angle of 168º, covering radii up to 6,000 km from Arrokoth. The MVIC instrument, which has better rejection of scattered sunlight than LORRI, was used in its panchromatic framing mode, with total exposure times of 30 seconds. Reduction and analysis followed methodologies used for similar Pluto data (*48*). No rings or dust structures were detected, with an upper limit I/F of ~1.5 x $10^{-6}$ for structures wider than about 10 km in Arrokoth's equatorial plane (Fig. S4). Any ring around Arrokoth is thus also fainter in forward scattering than Jupiter's main ring (I/F = 4 x $10^{-6}$ at this phase angle, (*47*)).

New Horizons' Student Dust Detector (SDC) instrument (*49*) detected no signals above the noise threshold within ± 5 days of the Arrokoth encounter, implying that there were no impacts by dust particles > 1.6 µm in radius, giving a 90% confidence upper limit of 3 x $10^7$ particles km$^{-2}$. For 10% albedo, this is equivalent to an I/F limit of 3 x $10^{-11}$, even more constraining than the optical limit, for particles of this size or larger along the spacecraft trajectory.

**Comparison to Other KBOs, and to Possible Captured KBOs**

Though most other known CCKBOs are larger than Arrokoth, due to observational biases, Arrokoth appears typical of CCKBOs using the few metrics that can be directly compared. Arrokoth's 0.6 µm geometric albedo, 0.23, is within the known range of other CCKBOs (*50*). Rotational lightcurves suggest that up to 25% of larger CCKBOs could be contact binaries like Arrokoth (*13*), though contact binaries appear to be more abundant, up to 50%, in the Plutino population (*51*). Arrokoth's color is also typical of CCKBOs (*1,3*).

Many irregular satellites of the giant planets may be captured KBOs, but only three have resolved spacecraft images. Neptune's satellite Triton, diameter 2700 km, is far too large and active to be a useful comparison body to Arrokoth. Neptune's smaller irregular satellite Nereid, 170 km in diameter, has a geometric albedo of 0.16 – 0.20, similar to Arrokoth, but is neutral in color (*52*). Saturn's 210 km diameter irregular satellite Phoebe (possibly a captured Kuiper Belt Object (*53*), though perhaps instead a captured C-type asteroid (*54,55*)), is darker (geometric albedo 0.08 (*56*)) and less red (*57*), and has a completely different surface appearance, dominated entirely by impact features (*58*). If Phoebe ever resembled Arrokoth, it has been drastically altered by subsequent evolution.

**Comparison to Jupiter Family Comets**
A class of objects previously explored by spacecraft that may be analogous to Arrokoth in ultimate origin are the Jupiter Family Comets (JFCs). These differ from Arrokoth in three major respects: (i) Provenance: the vast majority of these bodies likely originated in the Kuiper belt, but from a different family of KBOs: the population of "scattered KBOs" which likely originated closer to the sun than Arrokoth, and whose orbits are strongly perturbed by gravitational interactions with Neptune (*59*); (ii) Size: the effective spherical diameters of the JFC nuclei visited by spacecraft are 3 to 18 times smaller than that of Arrokoth; (iii) Thermal history: JFCs have experienced intense solar heating which has heavily modified their surfaces. By comparing the properties of Arrokoth and JFC nuclei, we can explore the effects of these differences.

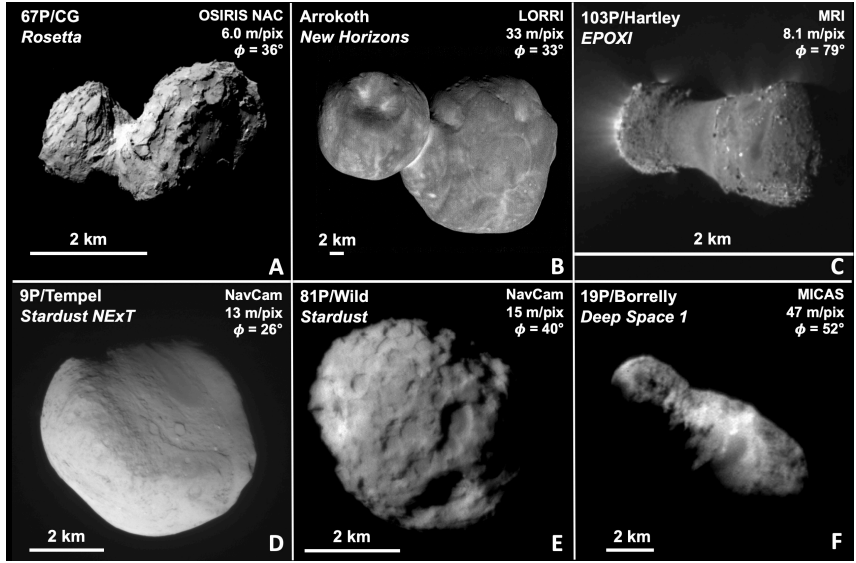

**Fig. 8. Comparison of JFC nuclei to Arrokoth.** The images of JFC nuclei have phase angles similar to those of the highest resolution image of Arrokoth, except for 103P, which was only observed at much higher phase angles. A: Rosetta image of 67P/Churyumov–Gerasimenko (*73*); B: New Horizons image of Arrokoth (this paper); C: Extrasolar Planet Observation and Characterization / Deep Impact Extended Investigation (EPOXI) image of 103P/Hartley (*74*); D: Stardust image of 9P/Tempel (*75*) E: Stardust image of 81P/Wild (*76*); F: Deep Space 1 image of 19P/Borrelly (*77, 78*, credit: NASA/JPL). Each frame is scaled so that the body nearly fills it, with the true relative sizes of each body indicated by the scale bars. Arrokoth is much larger than these comets. Figure S3 shows the equivalent images scaled to the same linear resolutions.

The JFC nuclei visited by spacecraft have diverse shapes and surfaces (Fig. 8, Fig. S3 and Table S3)). Comets 19P, 67P, and 103P appear to be highly elongated bilobate objects, suggesting the merger of two distinct bodies, as has been proposed for Arrokoth (*1,18*), though for comets it is also possible that thermal evolution has generated this shape (e.g., *60*). Except for 67P, whose bulk density is 538 ± 1 kg m$^{-3}$ (*16*), the densities of the other JFC nuclei are uncertain by a factor of two or more, but all are consistent with ~500 kg m$^{-3}$ (*61*), which implies average bulk porosities of ~50-80%. Arrokoth's density is likely greater than 290 kg m$^{-3}$ (see above), and thus at least consistent with those of JFC nuclei. The rotation period of Arrokoth is similar to those measured for 67P and 103P and falls well within the range measured for the JFC population (*62*), though JFC rotation is known to be affected by cometary activity (*63*).

The JFC nuclei listed in Table S3 are much darker than Arrokoth, with ~3 - 5 times smaller geometric albedos. If the JFC nuclei once had higher albedos in their nascent state in the Kuiper belt, then the darkening of their surfaces might be associated with cometary activity while the JFCs are in the inner Solar System. Most surface features on JFC nuclei have been attributed to cometary activity (e.g., *64, 65*). Generally, the surfaces of JFC nuclei can be divided into "smooth" and "rough" (or "mottled") regions, with the rough terrains associated with a preponderance of pits/depressions or mounds/hills (*66,67*). The smooth regions of JFCs are generally brighter than average and are often associated with topographic lows, suggesting accumulation by small grains that scatter light more efficiently than the average surface, as we proposed for Arrokoth above. However, on comets the fallback of grains ejected by sublimation is likely to be contribute to smooth terrains (*68*), and this is less likely to be important on Arrokoth where evidence for sublimation erosion is limited to the pit chains of possible sublimation origin, and tentative evidence for scarp retreat on the small lobe, as mentioned above.

While the large (multi-kilometer) scale bilobate morphology of Arrokoth is similar to 4 out of the 6 comets listed in Table S3 (see also Figs. 8, S3), the finer surface textures are not. JFCs imaged at the same resolution as Arrokoth show fewer impact craters than Arrokoth (*64*), consistent with these comets having highly erosional surfaces. They may lose their surfaces at ~0.5-1.0 m per orbit (*69*) with 5-10 yr orbital periods, so small pits will be removed within a few thousand years. They also show a much rougher surface texture at the 50-100 m scale, consistent with sublimation erosion and loss of most of the erosional debris.

**Conclusions**

Our dataset from the New Horizons flyby of Arrokoth provides a more complete picture of the physical nature of this object. Images taken on approach show that while both components of Arrokoth are flattened, the flattening is less extreme than initially inferred (*1*), and the two components have a larger volume ratio, $1.9 \pm 0.5$ than previous estimates. Stereo topography and the highest resolution imaging taken during the flyby show that the large lobe is very flat on the encounter hemisphere. If the large lobe is composed of multiple components which accreted separately, as previously proposed (*1*), the topographic signature of the boundaries between the components would be expected to be large initially, if the sub-units were mechanically similar to the two present lobes at the time of their coming into contact (*18*). The observed flatness of the large lobe shows that any such discontinuities have been subdued, and in some cases eliminated entirely. If subsequent deposition subdued the boundaries, post-depositional processes must be invoked to explain why many of the boundaries are still visible as differences in surface texture or as linear albedo features. Alternatively, the large lobe may be a monolithic body, and the apparent division into sub-units may be due entirely to secondary processes. Multiple processes, including impacts, have reworked the surfaces of both lobes after their formation, producing the fissures, small dark hills, and sinuous albedo boundaries seen in the images.

Crater densities on Arrokoth are low but consistent with a surface age of > 4 Ga, due to the expected low cratering rates in the CCKB, even if only craters with the highest confidence of being impact features are included in the counts. This dates the surface as plausibly from the end of Solar System accretion. Crater size-frequency distribution slopes for < 1 km craters on Arrokoth are poorly constrained, but are consistent with the slopes seen for 2 – 15 km craters in the Pluto system (*39*), suggesting that the shallow size-frequency distribution for 0.2 - 2 km diameter KBO impactors found by (*39*) may persist down to smaller sizes.

Arrokoth is unlike other small bodies visited by spacecraft. The surfaces of comets are dominated by volatile loss and sublimation erosion driven by the thermal energy inputs, due to their position in the inner Solar System. The surfaces of asteroids are dominated by high-energy impacts. As a result, asteroid surfaces are primarily rubble or impact ejecta. In both cases the dominant energy environment (thermal and impact) is driving the surface morphology. Arrokoth's surface is probably a consequence of its presence in the CCKB, where there is much less energy input. The very small relative velocities in this dynamical population result in few impacts and those that do occur have very slow impact velocities. Without strong energy inputs either from solar radiation or impacts, the surface of Arrokoth is expected to be dominated by low level energy inputs from interstellar, solar, and micro-meteorite energy sources at slow rates, likely extending to just a few meters depth (*27*). It is this low-energy environment that has allowed its surface to be preserved for four billion years.

Arrokoth appears to be a typical CCKBO, to the extent that we can compare it to others, so it can be used to understand the cold classical belt as a whole. The bi-lobed nature of Arrokoth might be common in the Kuiper Belt, and could indicate that the bi-lobed shape of many comet nuclei is a primordial feature. In addition, Arrokoth appears to be a direct product of accretion rather than a collisional fragment, and is much smaller than the ~100 km diameter of the break in slope of the size-frequency distribution of CCKBOs (*6,70*). These facts are consistent with the break in slope being a primordial feature, as predicted by streaming instability models (*71*). Arrokoth's appearance is much less consistent with the break in slope being a result of later destruction of small CCKBOs by collisions, a hypothesis also inconsistent with the observed deficit of small craters in the Pluto system (*39*).

**Acknowledgments:** We thank all who contributed to the success of the New Horizons flyby of Arrokoth, and in particular the National Astronomical Observatory of Japan's Subaru Telescope, the Carnegie Observatory's Magellan Telescopes, the Canada-France-Hawaii Telescope, the NASA Hubble Space Telescope, the Harvard-Smithsonian Center for Astrophysics, the Massachusetts Institute of Technology, Northern Arizona University, the University of Hawaii, the Hertzberg Institute for Astrophysics, and NASA, for their support of the search campaign that led to its discovery. We also are indebted to the Hubble Space Telescope and the European Space Agency's Gaia mission for their key roles in the precise orbit determination required to enable the successful flyby. **Funding:** Supported by NASA's New Horizons project under contracts NASW-02008 and NAS5- 97271/TaskOrder30. Also supported by the National

# Supplementary Materials

**Materials and Methods**

Image Processing

Each set of LORRI images of Arrokoth consists of multiple consecutive frames.  In the case of the closest approach ("CA") image sets (Table S1), the images were taken during simultaneous scans by the Ralph instrument, and thus include some motion smear.  For each image set, frames are registered by shifting, scaling, and rotating as necessary, and are stacked to improve the SNR.  This stacking process removes features such as image defects and background stars, which are not present at the same location relative to Arrokoth in every frame. Finally, images are deconvolved to remove much of the LORRI point-spread function, and also to remove motion smear in the case of the CA image sets, to produce the final images used for analysis.

Construction of Stereo Models

Three sets of LORRI images, CA04, CA05, and CA06 (Table S1) provide the highest-resolution stereo coverage.  The CA05 / CA06 pair has the best nominal resolution and a stereo convergence angle of 17°.  However, the CA04 / CA06 pair (Fig. 1A) provides better stereo because, in addition to a slightly larger convergence angle (almost 20°), CA04 has a much longer effective exposure time (thus better SNR), and also lower smear, than CA05 (Table S1).  We used the Ames Stereo Pipeline (*80,81*) on the stacked, deconvolved products from the CA04 & CA06 observations to derive a stereographic terrain model of the surface of Arrokoth.  An iterative closest point algorithm (*82,83*) was used to rigidly rotate and translate the stereo model surface to match the -Z facing surface of the global shape model, though the required rotation was < 0.5º

Geomorphological Mapping of Arrokoth

Constructing a planetary geomorphological map derived solely from images acquired above the study area involves defining and characterizing discrete material units based primarily on the surface morphology, texture, albedo, and color as seen at the pixel scale, which are physical attributes that are related to the geologic processes that produced them. Along with visible structural features, the distributions of these units are then mapped to identify the relative roles of different geological processes that shape planetary surfaces. We have followed standard US Geological Survey mapping protocol (*84*) when creating our geomorphological map of Arrokoth (Fig. 1C), although applying the principles of mapping to it can be challenging, primarily because our highest resolution observations of the target (138 to 33 m pixel$^{-1}$) were only obtained at relatively low phase angles (12.9° to 32.5°). Outside a narrow strip near the terminator, the low phase angle hinders assessment of topography at a scale of hundreds of meters based on surface shading. In addition, the consistently low phase of the approach imaging generates uncertainty regarding how much of the observed surface heterogeneity across Arrokoth is due to intrinsic geological variation, or is a consequence of variable illumination of a limited range of geomorphological units. We have created a geomorphological, rather than a geological map, i.e. the units that we have defined for Arrokoth have been inferred from what appear to be distinct physiographic components of the two lobes, but the map is not intended to rigorously convey stratigraphic relations between units. Stratigraphic organization of the units would require application of the rules of superposition and crosscutting, which we do not consider to be feasible given the limitations of available data and inherent ambiguities associated with its interpretation. Instead, the map is intended to reduce the complexity of Arrokoth's surface to comprehensible proportions that are more amenable to the development of hypotheses for the formation and evolution of Arrokoth.

    On the large lobe, which shows less overall albedo variation but more limb topographic amplitude than the small lobe, the boundaries of the individual sub-units that compose the lobe have been defined based largely on topographic expression in the LORRI CA06 33 m pixel$^{-1}$ imaging, and in stereo imaging (e.g. comparing LORRI CA04 and CA06 in Fig. 1A). High solar incidence angles near the terminator make troughs and scarps separating sub-units visible in this region (including those separating units *ta, tb, tc,* and *sm*). In contrast, the boundary of unit *tg*, located on the limb of the large lobe, is inferred not based on shading due to topography, but due to it being ringed by bright material (unit *bm*), which we interpret to be loosely-consolidated material that has collected in depressions across Arrokoth, particularly at the neck connecting the two lobes. Stereo imaging is necessary for the identification of unit *tf* as a separate unit, as its apparent position can be seen to move relative to unit *te* between LORRI CA04 and CA06. While albedo variation across the large lobe is less than on the small lobe, some units are distinguished by their albedo characteristics, such as units *ta, tc,* and *td*, which appear lighter-toned than the neighboring units of *te, th,* and *sm*, despite these latter units being illuminated at lower solar incidence angles. Unit *tg* is distinguished from unit *te* by the partial stretch of bright material that exists between them, and because unit *tg* displays a surface pattern characterized by albedo contrasts on a scale of hundreds of meters, whereas unit *te* appears dark and homogeneous at this scale. The boundary between units *sm* and *th* is located in the center of the face of the large lobe, far from the terminator and the limb, and there is no apparent topographic discontinuity associated with it. Instead, a tentative contact has been defined based on the differing textures presented by these units (unit *th* shows greater albedo contrast than unit *sm*) as well as the presence of a portion of the bright annulus that separates them, although we have identified locations where some darker elements of unit *th*, apparently hills, extend across the annulus. We treat the distinct physiographic units on the large lobe as individual

geomorphological units, although it is possible that they are in fact all topographic expressions of the same unit.

On the small lobe, units are defined primarily according to the different albedos and planforms they present; any topographic signatures associated with them are much less apparent when compared with those of sub-units under similar lighting conditions on the large lobe, which is at least in part due to the smaller scales of the small lobe's units. The small lobe's limb topography, however, does indicate a break in slope that corresponds to the stretch of dark material (unit *dm*) that separates units *mm* and *rm*, and suggests that unit *mm* occupies a local high, whereas unit *rm* occupies a local low. This topographic discontinuity is an important factor in the decision to map these areas as separate units: whereas unit *rm* displays a pitted surface and unit *mm* does not, the two units cannot be mapped separately based on this criterion alone, as we cannot rule out that variable illumination has played a role in contributing to their different appearances, given the more oblique lighting of unit *rm*.

These examples demonstrate how we considered every aspect of available imaging (in particular stereo parallax, surface shading at low and high solar incidence angles, and limb topography to identify discrete geomorphological units given the limited data available.

The R Value Measure of Crater Densities

The R value plotted on Fig. 6B is constructed from a differential power law size-frequency distribution of crater diameters ($dN/dD \propto D^q$) normalized by a $D^{-3}$ distribution, where $N$ is the crater spatial density and $D$ is the diameter. The power law exponent ($q$) is commonly referred to as the distribution log-log slope. In this visualization, a crater size-frequency distribution with a slope $q$ of -3 appears as a horizontal line, allowing differences from the common $D^{-3}$ distribution to be easily seen.

Crater Identification and Classification

Feature size measurements were carried out on the original deconvolved CA06 LORRI image (Fig 1A). As no projection was used, the crater sizes were measured in pixels and converted to approximate sizes using the pixel scale of 33 m px$^{-1}$. We measured the most representative diameter of a feature (e.g., measuring the long axis of obliquely viewed features to avoid foreshortening where possible). The size for the crater Maryland is an average of 6 chords. Some features have more distinct boundaries than others. For interpreting crater diameters given here, we use a diameter uncertainty, tabulated in Data S3, of 2 pixels or 20% of the crater diameter, whichever is larger for a given feature (i.e., for feature diameters above 10 pixels, or ~330 m, we use 20%).

Data S3 includes the crater sizes and subgroup designations for each feature considered for crater analysis and shown in Fig. 6A. Some features were determined to have a low likelihood of being either fresh or modified impact craters, and thus are not included in any subgroup plotted in Fig. 6B. The descriptions for each subset are:

- The A_High (Arrokoth high confidence) subgroup includes only features 0.34 km (~10 pixels) or larger in diameter, only fairly circular features, and features with the topography expected of impact craters. This subset includes a few features that are more subtle or shallow than the deepest probable craters on Arrokoth, but they are all close to the terminator where the low sun angle makes clear their likely identification as impact craters.
- The A_Medium (Arrokoth medium confidence) subgroup includes smaller and/or less circular features. It includes 4 features less than 0.34 km in diameter, from ~0.23 – 0.27 km across.

- The A_Low (Arrokoth low confidence) subgroup includes features that are depressions or bright spots but considerably less circular, and features in a chain that may be associated with a tectonic feature (at a subunit boundary). It includes 6 features less than 0.34 km in diameter, from ~0.19 – 0.28 km across.
- The LL_Bright (Large lobe bright spot) subgroup, designed to give an approximation of a maximum density, includes: all A_High, A_Medium, and A_Low features larger than 0.27 km (8 pixels), that are bright, circular or sub-circular features on the sunward half of the large lobe only (right of the solid line in Fig. 6A).
- The A_Pits (Arrokoth pits) subgroup, designed to give an approximation of a maximum density, includes all A_High, A_Medium, and A_Low features larger than 0.27 km, both circular and sub-circular, and also includes few features in a chain, that are on the anti-sunward half of the large lobe only (left of the solid line in Fig. 6A).
- The LL_Term (Large lobe terminator) subgroup, designed to give an approximation of a most likely density, includes: all A_High and A_Medium features larger than 0.27 km, both circular and sub-circular, in the near-terminator region left of the dashed line on Fig. 6A.

Additional information is in Table S2.

Details of Satellite Searches

Searches conducted to assess flyby hazards from 42 to 19 days before the flyby covered the entire Hill sphere (~40,000 km radius assuming an Arrokoth density of 500 kg m$^{-3}$) with a range of total exposure times up to 1 hour. Later 2×2 and 4×4 frame image mosaics, taken from 3.3 days to 6 hours before the flyby with 2.6 – 12 minute total exposures, covered smaller regions with greater sensitivity. Each search consisted of 2 – 3 mosaics taken 0.6 – 2.0 hours apart, to identify satellites by their motion relative to the dense Milky Way star background. These deep searches overexposed Arrokoth and thus had limited sensitivity very close to it, so close approach images exposed for Arrokoth's surface were used to search for satellites with very small orbital radii. Sensitivity limits were established by implanting synthetic objects into the original images.

**Supplementary Text**

Expected Impact Crater Morphologies

Despite the low impact velocities, we expect most impacts on Arrokoth to form craters similar to those seen elsewhere in the Solar System. Crater morphology varies with the impactor and target characteristics. The low and high velocity tails of the expected impact velocity distribution for Arrokoth reach down to a few m s$^{-1}$ and as high as a few km s$^{-1}$, but the mode, ~300 m s$^{-1}$ (*5*), is slow compared to primary cratering velocities on the surfaces of both icy and rocky bodies closer to the Sun, and is more typical of secondary cratering velocities on those bodies (*85, 86*). These impacts often form craters with similar morphological characteristics to primary craters, although secondary craters are often shallower than the same size primary impact, and may be elongated in the direction radial to the primary crater. We do not suggest that any craters on Arrokoth are secondary craters, but secondary craters elsewhere show that 300 m s$^{-1}$ impacts are capable of creating craters on the surface of Arrokoth. The formation of a crater on a slope or modification by later geologic processes (such as mass wasting or a subsequent fault near the crater) may also alter the crater's appearance.

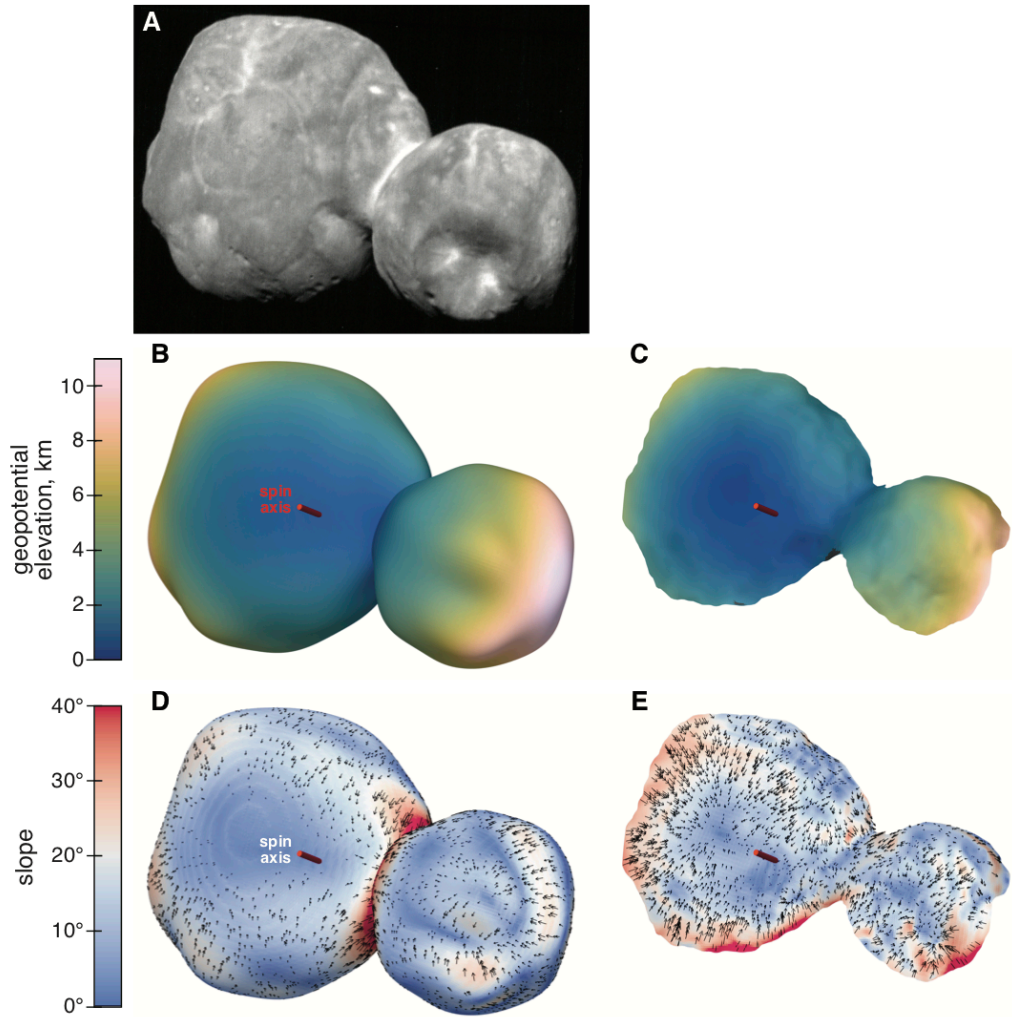

**Fig. S1. Slopes and Gravity of Arrokoth.** CA06 image of Arrokoth (A) compared to illustrations of gravitational parameters seen from the same geometry. B and C: Geopotential elevation for the global shape model (B) and stereo model (C). D and E: Slopes computed from the global shape model (D) and stereo model (E), for an assumed density of 500 kg m$^{-3}$. Color gives slope magnitude, and arrows give the slope direction.

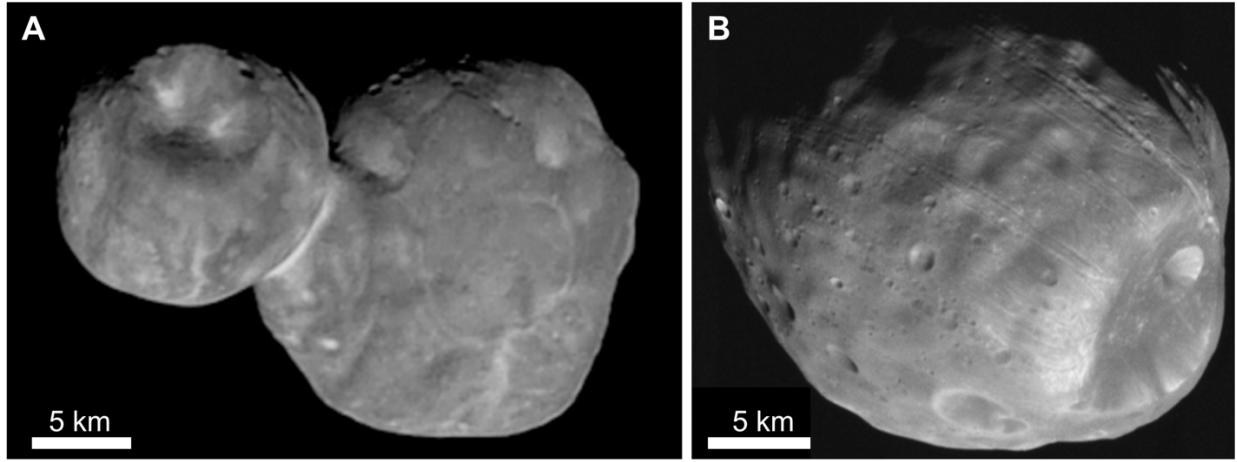

**Fig. S2. Craters on Arrokoth compared to those on the Martian moon Phobos.** A: Our highest resolution Arrokoth image (CA06) at 32° phase. B: An image of the Martian moon Phobos (right, diameter = 22.5 km) from the Mars Reconnaissance Orbiter, obtained at a similar but slightly lower phase angle (26.4°) (Credit: NASA/JPL-Caltech/University of Arizona). The image of Phobos has been processed to match the pixel scale, smear, camera point-spread-function, SNR, and deconvolution of our highest resolution LORRI images of Arrokoth (*87*). Many more unambiguous craters can be seen across the surface of Phobos than on Arrokoth.

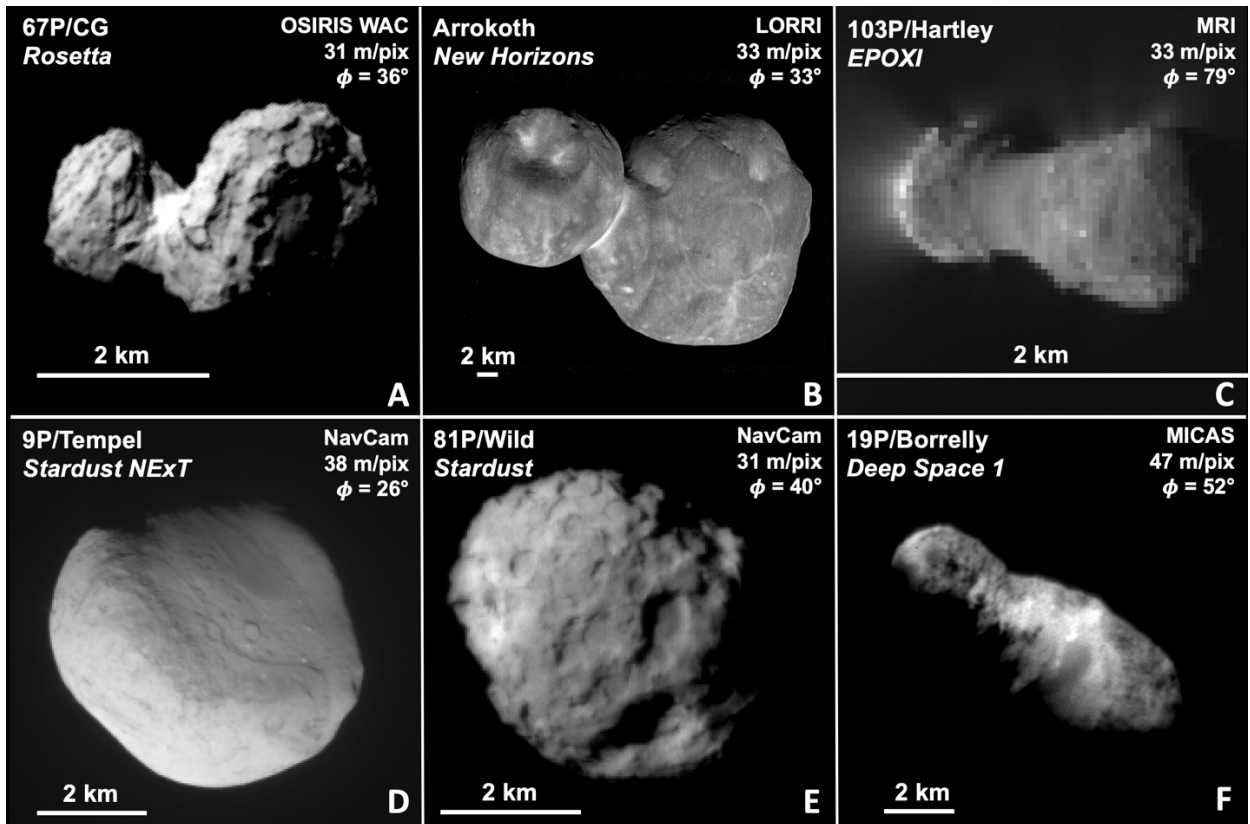

**Fig. S3. Images of JFC nuclei and Arrokoth at comparable pixel scale.** A: Rosetta (*73*); B: New Horizons (this paper); C: EPOXI (*74*); D and E: Stardust (*75,76*); F: Deep Space 1 (*77*). For higher-resolution image comparisons, and additional details, see Fig. 8.

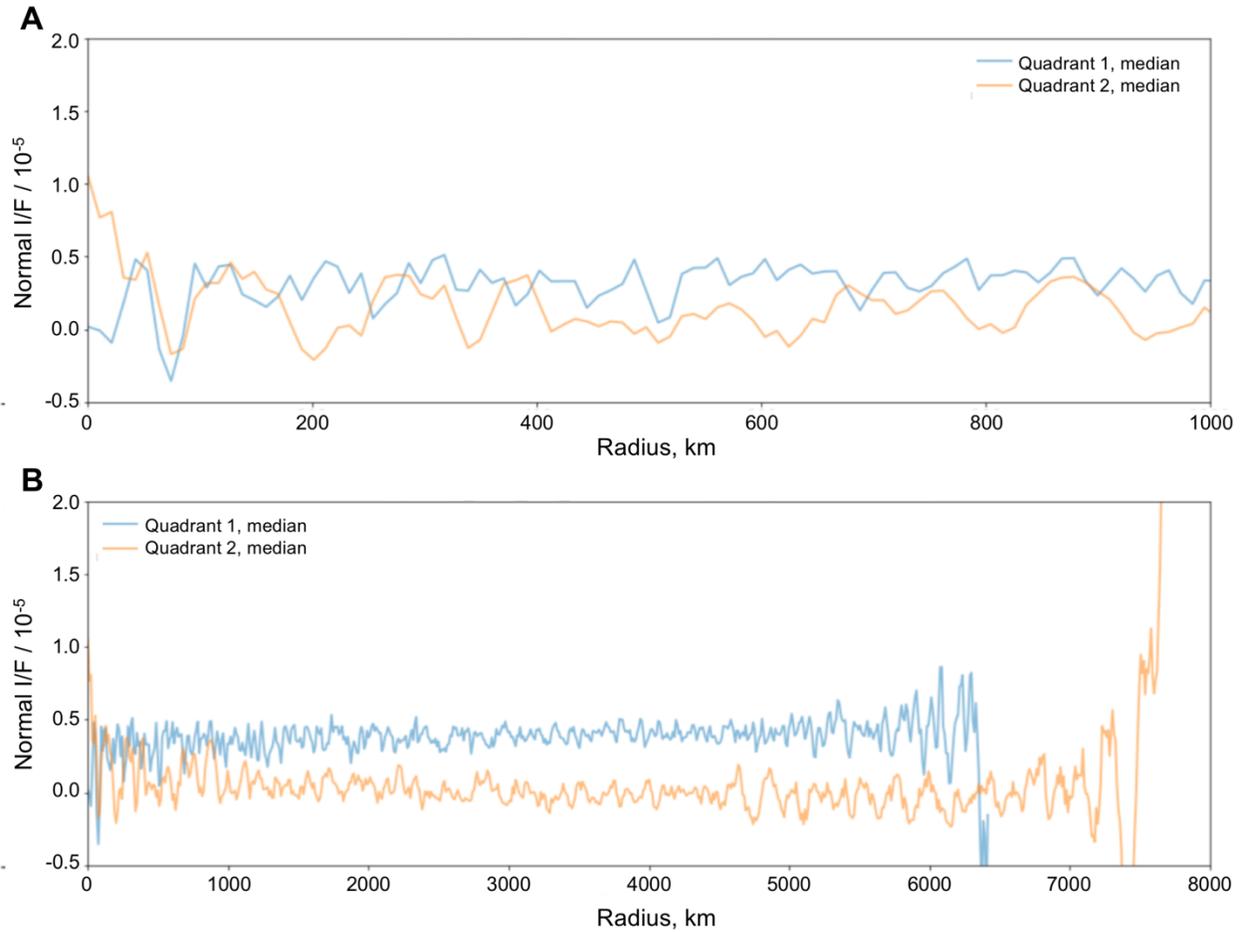

**Fig. S4. Non-Detection of forward-scattering rings around Arrokoth**. Radial profiles of the sky brightness, in units of I/F (*19*), as a function of distance from Arrokoth in its equatorial plane, from MVIC images taken 1.7 – 2.3 hours after closest approach at a phase-angle of 168º. A: The innermost region. B: The entire profile. The blue and orange curves were derived from different quadrants of the image: the vertical offset between them is an artifact. Profiles are binned to a radial resolution of 10.6 km. No rings or dust structures were seen, with an upper limit I/F of ~$1.5 \times 10^{-6}$ for structures wider than about 10 km.

**Table S1. Close approach LORRI images**. The image set name "CAnn" refers to the n[th] observation of the close approach (C/A) sequence. CA03 and CA08, not listed here, were radiometric, not imaging, observations.

| Image set name | Mode | Mid-Time, mins. after C/A | Range, km | Phase, degrees | Resolution, km pixel$^{-1}$ | Smear, pixels | Single-Frame Exposure Time, sec | Number of Co-added Frames | Combined Exposure Time, sec |
|---|---|---|---|---|---|---|---|---|---|
| CA01 | LORRI 1×1 | -70.6 | 61,214 | 11.8 | 0.304 | 0.6 | 0.150 | 43 | 6.45 |
| CA02 | LORRI 1×1 | -49.1 | 42,663 | 12.0 | 0.212 | 4.0 | 0.025 | 6 | 0.15 |
| CA04 | LORRI 1×1 | -31.9 | 27,850 | 12.9 | 0.138 | 0.4 | 0.100 | 25 | 2.50 |
| CA05 | LORRI 1×1 | -18.8 | 16,680 | 15.7 | 0.083 | 4.0 | 0.025 | 6 | 0.15 |
| CA06 | LORRI 1×1 | -6.5 | 6,634 | 32.5 | 0.033 | 4.0 | 0.025 | 6 | 0.15 |
| CA07 | LORRI 4×4 | 9.4 | 8,834 | 152.4 | 0.175 | 8.1 | 0.200 | 6 | 1.20 |

**Table S2. Feature subgroups for crater analysis**. Data S3 includes the full list of craters and sizes with classification information. The diameter range column does not include Maryland (6.7 km diameter), which is treated separately.

| Subgroup | Lobe where Present | Number of features | Diameter range (km) | Surface area used (km²) | Area description |
|---|---|---|---|---|---|
| A_High | Both lobes | 10 | 0.34–7.16 | 700 | Entire visible surface of Arrokoth: Area is ½ of the global shape model surface area |
| A_Medium | Both lobes | 17 | 0.24–0.64 | | |
| A_Low | Both lobes | 16 | 0.19–0.68 | | |
| LL_Bright | Large lobe only | 10 | 0.27–0.62 | 230 | Sunward half of large lobe: Area is ½ of large lobe's visible surface area |
| LL_Pits | Large lobe only | 15 | 0.27–0.77 | 230 | Anti-sunward half of large lobe: Area is ½ of large lobe's visible surface area |
| LL_Term | Large lobe only | 7 | 0.27–0.77 | 90 | Measured from the shape model for the selected area |

**Table S3**. **Properties of Arrokoth compared to cometary nuclei**. The object IDs are listed in the first column. The second column refers to the best fit ellipsoid dimensions, even though the actual shape may differ significantly from an ellipsoid. The effective spherical diameters, calculated from the best fit ellipsoidal dimensions in the second column, are presented in the third column and provide perhaps the best single number for the size of the object. The density for 67P is exceedingly well determined (*16*) and lies roughly in the middle of the ranges estimated for JFC nuclei. Due to the jetting force from cometary outgassing, the rotational periods of JFC nuclei change with time, so only approximate current values are listed for them. The JFC geometric albedos are for a wavelength of 550 nm (V-band) and are taken from (*88,89*), sometimes with small corrections to transform from R-band (650 nm) to V-band using the typical value for JFC colors as reported in (*90*). The geometric albedo for Arrokoth (this work) is for a wavelength of 600 nm. The variation of reflectance across the surfaces of the JFC nuclei and Arrokoth are comparable (±15-20% variation about the global mean value), except for 19P, which shows a variation about twice that of the other objects, apparently associated with two different types of terrains (*78*).

| Object ID | Ellipsoid Axes (km) | Spherical Diameter (km) | Density (kg m$^{-3}$) | Rotational Period (hr) | Geometric Albedo |
|---|---|---|---|---|---|
| Arrokoth | 36 × 20 × 10 | 18.3 | >290 | 15.9 | 0.23 |
| 9P/Tempel | 7.6 × 4.9 × 4.6 | 5.6 | 200-600 | ~41 | 0.056 |
| 19P/Borrelly | 8.0 × 3.2 × 3.2 | 4.3 | 290-830 | ~25 | 0.065 |
| 67P/Churyumov-Gerasimenko | 4.3 × 2.6 × 2.1 | 2.9 | 538 ± 1 | ~12 | 0.058 |
| 81P/Wild | 5.5 × 4.0 × 3.3 | 4.2 | - | - | 0.059 |
| 103P/Hartley | 2.2 × 0.5 × 0.5 | 0.92 | 200-400 | ~18 | 0.045 |